\documentclass[preprint,aps,showpacs,preprintnumbers,amsmath,amssymb,
nofootinbib]{revtex4}
\usepackage{epsfig}
\begin{document}

\begin{flushright}
\end{flushright}
\thispagestyle{empty}

\newcommand{\be}{\begin{equation}}
\newcommand{\ee}{\end{equation}}
\newcommand{\bea}{\begin{eqnarray}}
\newcommand{\eea}{\end{eqnarray}}
\newcommand{\bers}{\begin{eqnarray*}}
\newcommand{\eers}{\end{eqnarray*}}
\newcommand{\nn}{\nonumber}
\newcommand\un{\cal{U}}
\def\dis{\displaystyle}
\def\B{B_d^0}
\def\Bb{\bar{B}_d^0}
\def \drho{\bar \rho}
\def \deta{\bar \eta}
\def\l{\lambda}
\def\s12{s_{12}}
\def\c12{c_{12}}
\def\s13{s_{13}}
\def\c13{c_{13}}
\def\s23{s_{23}}
\def\c23{c_{23}}
\def\ed{e^{i \delta}}
\def\edm{e^{-i \delta}}


\title{\large Probing CP violation in the neutrino sector with magic
baseline experiments }
\author{ Rukmani Mohanta$^1$ and Daughty John$^2$  }
\affiliation{$^1$ School of Physics, University of Hyderabad,
Hyderabad - 500 046, India} \affiliation{$^2$ Department of Physics,
Indian Institute of Technology Hyderabad, ODF Estate, Yedumailaram -
502205, Andhra Pradesh, India}

\begin{abstract}

We investigate the effect of CP violation in the leptonic sector.
Due to the tiny neutrino masses  its value is predicted to be very
small and it is far beyond the experimental reach of the current
experiments. Recently, the magic baseline experiment from CERN to
INO (Indian Neutrino Observatory) with $L=7152$ km has been proposed
to get a sensitive limit on $\sin \theta_{13}$. We show that due to
such magic baseline neutrino beam it is possible to observe CP
violation in the neutrino sector upto several percent for the beam
energy between (1-10) GeV.

\end{abstract}
\pacs{14.60.Pq, 11.30.Er} \maketitle

It is now well established by the recent neutrino oscillation
experiments \cite{ref1a, ref1b,ref1c,ref1d,ref1e,ref1f,ref1g,ref1h}
that neutrinos do have a tiny but finite nonzero mass. Because of
the non-zero mass, the flavor eigenstates of the neutrinos are no
longer be the corresponding mass eigenstates and these two are
related  by some unitary transformation. Thus, due to the mixing
between the flavor and mass eigenstates  of neutrinos, it is
expected that there could also be $CP$ violation in the neutrino
sector analogous to that of the quark sector.  CP violation so far
has been observed only in the quark sector of the standard model
i.e., in the $K$ and $B$ meson systems, the origin of which is
basically attributed to the complex phase in the
Cabibbo-Kobayashi-Maskawa (CKM) mixing matrix \cite{ckm,ckm1}. Its
discovery in the leptonic sector should shed additional light on the
understanding of the origin of CP violation in nature. The study of
CP violation in the lepton sector though less examined than that of
the quark sector, it is indispensable, since neutrinos are allowed
to be massive and the corresponding mixing matrix is complex. It
seems necessary for us to examine whether there is a chance  to
observe CP violation in the leptonic sector in the long baseline
experiments. In this paper we explore such a possibility.

 Let us briefly review the CP violation phenomenon in
neutrino oscillation experiments  to clarify our notation. Within
the framework of three lepton families, the three flavor eigenstates
of neutrinos ($\nu_e,~ \nu_\mu,~ \nu_\tau$) are related to the
corresponding mass eigenstates ($\nu_1,~ \nu_2,~ \nu_3$) by the
unitary transformation

\bea
 \left ( \begin{matrix}
\nu_e       \\ \nu_\mu \\ \nu_\tau \\ \end{matrix} \right ) \;
=\;U\left ( \begin{matrix}  \nu_1 \cr \nu_2 \cr \nu_3
\cr\end{matrix} \right )~\equiv~ \left ( \begin{matrix} U_{e1}
& U_{e2} & U_{e3} \\ U_{\mu 1} & U_{\mu 2}    & U_{\mu 3} \\ U_{\tau
1} & U_{\tau 2}    & U_{\tau 3} \\ \end{matrix} \right ) \left (
\begin{matrix}  \nu_1 \cr \nu_2 \cr \nu_3 \cr\end{matrix} \right )
\; ,\label{eq}
\eea where $U$ is the $3 \times 3 $ unitary matrix known as PMNS
matrix \cite{pmns, pmns1}, which contains three mixing angles and
three CP violating phases (one Dirac type and two Majorana type).
The unitary matrix $U$ can be represented in the standard
parametrization \cite{pdg}
 as \bea
U &= & \left ( \begin{matrix} 1 & 0& 0\\
            0 & c_{23} & s_{23} \\
0 & -s_{23}& c_{23}\\
           \end{matrix} \right )\left ( \begin{matrix}
c_{13} & 0& s_{13} e^{-i \delta}\\
            0 & 1& 0 \\
-s_{13} e^{i \delta} & 0 & c_{13}\\
           \end{matrix} \right )\left ( \begin{matrix} c_{12} & s_{12}& 0\\
            -s_{12} & c_{12} & 0 \\
0 & 0 & 1\\
           \end{matrix} \right )
           \left ( \begin{matrix} 1 & 0& 0\\
            0 & e^{i \alpha} & 0 \\
0 & 0 & e^{i \beta}\\
           \end{matrix} \right )\nn\\
           &=& \left ( \begin{matrix} c_{12} c_{13} & s_{12}c_{13} &
           s_{13} \edm \\
-s_{12}\c23 - c_{12} s_{23} s_{13} e^{i \delta } & c_{12} c_{23} -
s_{12} s_{23} s_{13} e^{i \delta} &
s_{23} c_{13} \\
s_{12} s_{23} - c_{12} c_{23} s_{13} e^{i \delta } & -c_{12} s_{23}
- s_{12} c_{23} s_{13} e^{i \delta }&
c_{23} c_{13} \\
\end{matrix} \right )  \left ( \begin{matrix} 1 & 0& 0\\
            0 & e^{i \alpha} & 0 \\
0 & 0 & e^{i \beta}\\
           \end{matrix} \right )\eea with $c_{ij}= \cos \theta_{ij}$, $s_{ij}=\sin \theta_{ij}$
and $\theta_{12},~ \theta_{23}$ and $\theta_{13}$ the three neutrino
mixing angles, $\delta$ is the Dirac type CP violating phase and
$\alpha$ and $\beta $ are Majorana phases.  The presence of the
leptonic mixing, analogous to that of quark mixing, has opened up
the possibility that CP violation could also be there in the lepton
sector as it exists in the quark sector. Although the absolute
masses of the neutrinos are not yet known, the recent  experiments
like SNO, KamLand, K2K and MINOS \cite{ref1a,
ref1b,ref1c,ref1d,ref1e,ref1f,ref1g,ref1h, kam, kam1, k2k, minos}
provide information on the two mass square differences $\Delta
m_{21}^2$ and $\Delta m_{31}^2$ and on the two mixing angles
$\theta_{12}$ and $\theta_{23}$. The third mixing angle
$\theta_{13}$ is not yet determined but from the null result of
CHOOZ \cite{chooz} experiment, its value is expected to be quite
small. The current best fit values with $1 \sigma$ errors for three
flavour neutrino oscillation parameters from global fit \cite{best}
are given as \bea &&\Delta m_{21}^2 = \left ( 7.65_{-0.20}^{+0.23}
\right ) \times 10^{-5} ~{\rm eV^2},~~~~~~~\sin^2 \theta_{12}
=0.304_{-0.016}^{+0.022~},\nn\\
&&|\Delta m_{31}^2| = \left ( 2.40_{-0.11}^{+0.12} \right ) \times
10^{-3} ~{\rm eV^2},~~~~~~ \sin^2\theta_{23}
=0.50_{-0.06}^{+0.07}~,\nn\\
&&\sin^2\theta_{13}=0.01_{-0.011}^{+0.016}~,~~~(\sin^2\theta_{13}<0.04
~~({\rm 2 \sigma ~bound})),~~~~ \delta \in [0, 2
\pi],\label{data}\eea while the sign of $\Delta m_{31}^2$ is
unconstrained. The Majorana phases $\alpha$ and $\beta$ are
currently completely unconstrained.

 Let us take a closer look at the discovery
reach for CP violation. For this purpose we will first consider the
neutrino oscillation phenomenon in vacuum. From eq. (\ref{eq}), one
can write the evolution equation for the flavour eigenstates  as
 \bea i \frac{d}{dx} \nu_\alpha & = & -\left (U~{\rm
diag}(p_1, p_2, p_3)
~U^\dagger \right ) \nu_\alpha \nn\\
&\simeq &  \left ( -p_1 + \frac{1}{2E} U~ {\rm diag}(0, \Delta
m_{21}^2, \Delta m_{31}^2)~ U^\dagger \right ) \nu_\alpha\nn\\
& \simeq & \frac{1}{2E}\left ( U ~{\rm diag}(0, \Delta m_{21}^2,
\Delta m_{31}^2 )~ U^\dagger \right )\nu_\alpha, \label{eq1}\eea
where $p_i$'s are the momenta of the i'th-type mass eigenstates, $E$
is the energy and $\Delta m_{ij}^2=(m_i^2-m_j^2$) denote the
neutrino mass square differences. A term proportional to the unit
matrix like $p_1$ in eq. (\ref{eq1}) has been dropped because it is
irrelevant to the transition probability. The solution of
(\ref{eq1}) is given as \be \nu_\alpha (x)= U~\exp \left ( -i
\frac{x}{2E} {\rm diag} (0, \Delta m_{21}^2, \Delta m_{31}^2) \right
) U^\dagger ~\nu_\alpha(0). \ee

Thus, one can obtain the conversion probability for  $ \nu_\alpha
\to \nu_\beta $ process  at a distance $L$ as \bea P(\nu_\alpha \to
\nu_\beta;L) &=& \left | \sum_{i,j} U_{\beta i}\left [ \exp \left (
-i \frac{L}{2E} {\rm diag}(0, \Delta m_{21}^2, \Delta m_{31}^2)
\right )
\right]_{ij} U_{\alpha j}^* \right |^2\nn\\
&=& \sum_{i,j} U_{\beta i} U_{\beta j}^* U_{\alpha i}^* U_{\alpha j}
\exp\left (-i~ \Delta m_{ij}^2 (L/2E) \right ). \eea The simplest
measure of CP violation, which is equivalent to T violation if CPT
is conserved, would be the difference of oscillation probabilities
between neutrinos and antineutrinos, i.e., $P(\nu_\alpha \to
\nu_\beta)$ and $P(\bar \nu_\alpha - \bar \nu_\beta)$, which is
represented as \be \Delta P \equiv P(\nu_\alpha \to
\nu_\beta)-P(\bar \nu_\alpha - \bar \nu_\beta).\ee The transition
probability for the corresponding CP conjugate process $P(\bar
\nu_\alpha \to \bar \nu_\beta )$ can be obtained by replacing the
PMNS matrix elements $U_{\alpha i}$ by $U_{\alpha i}^*$. Thus, one
can obtain the CP or T violation parameter for the neutrino
oscillation case as
 \bea
\Delta P &\equiv &  P(\nu_\alpha \to \nu_\beta; L)-P(\nu_\beta \to
\nu_\alpha; L)\nn\\
&= &-4~ {\rm Im} (U_{\beta 1} U_{\beta 2}^* U_{\alpha 1}^* U_{\alpha
2}) (\sin 2 \Delta_{21} + \sin 2\Delta_{32}L+ \sin
2 \Delta_{13}  L)\nn\\
&=& 4 J f  \label{cpv}\eea where $\Delta_{ij} = \Delta m_{ij}^2 L/4E
$, and $L$ is the distance between the neutrino source and the
detector. $J$, the leptonic analog of Jarlskog Invariant and $f$ are
defined by \bea J&=& - {\rm Im} (U_{\beta 1} U_{\beta 2}^* U_{\alpha
1}^*U_{\alpha 2})
\nn\\
f &=& \sin 2 \Delta_{21} + \sin  2 \Delta_{32}  + \sin  2 \Delta_{13}\nn\\
&=& 4 \sin \Delta_{21} ~\sin \Delta_{32} ~ \sin \Delta_{13}  \eea

The size of $ \Delta P$ is proportional to $J$ times the product of
the sine of three mass differences. The effect is proportional to
$E^{-3}$ for small $\Delta_{ij}$. Therefore, there is a hope that
this effect will be visible in long baseline neutrino oscillation
experiment provided the Jarlskog invariant factor $J$ is not too
small.

In the standard parametrization of the mixing matrix \cite{pdg}, the
Jarlskog invariant  $J$ can be written as \bea J= {\rm Im}\left (
U_{\mu 3} U_{\tau 3}^* U_{\mu 2}^* U_{\tau 2} \right
)=s_{12}s_{23}s_{13}c_{12}c_{23}c_{13}^2\sin \delta \eea where
$\theta_{12}$ is the mixing angle that directly comes from solar
neutrino oscillation, $\theta_{23}$ is that for the atmospheric
neutrino oscillation  and $\theta_{13}$ is directly constrained by
the $\bar \nu_e \to \bar \nu_\tau$ oscillation experiment. Now using
the data from Eq. (\ref{data}), one can obtain the maximum value of
$J$ is given by \be J \leq 0.04~ \sin \delta .\ee Thus it is found
that the value of $J$ in the lepton sector is significantly larger
than that of the quark sector ($J_{quark} \sim{\cal O}(10^{-5})$),
provided $\delta$ is not too small.

The CP violation search will require pure neutrino beams with the
highest possible intensities. Beta-beams is a new concept for the
production of neutrino beams that is based on the beta-decay of
boosted radioactive ions, as first proposed by Zucchelli
\cite{Zucchelli:2002sa}. By exploiting the high ion intensities
foreseen in the future, this method can produce intense neutrino
beams, pure in flavour and with well known fluxes. The beta-beam
concept has several important advantages. The neutrino beams are
pure in flavour since only electron neutrinos or anti-neutrinos can
be produced, depending on the ion that decays through $ \beta^+$ or
$\beta^-$. This means that there is no beam related background. The
neutrino intensity and energy spectrum is precisely known, since the
number of ions is perfectly controlled.

In the standard beta-beam scenario \cite{Zucchelli:2002sa},  the
beta-beam facility is hosted at CERN. The search for CP violation
effects can be performed through the comparison of $ \nu_e \to
\nu_\mu$ versus $\bar \nu_e \to \bar \nu_\mu$ oscillations. If such
a beam is allowed to be detected at the INO detector, then the beam
has to travel a distance of 7152 km \cite{magic}, which is very
close to the magic baseline length $L_{magic}$ = (7300-7600) km
\cite{magic1, magic2}. At such a distance the $ \nu_e \to \nu_\mu$
survival probability has no dependence on $\delta$ and it allows to
measure the neutrino hierarchy without any degenerate solution. The
INO facility is expected to come up at PUSHEP situated close to
Bangalore at Southern India. It will have an Iron calorimeter (ICAL)
detector, which is expected to detect the charged muon with energies
of few GeV.

Now let us consider the matter effect in the CP violating parameter.
When the neutrino beam is allowed to travel a long distance, the
electron neutrinos  could have interaction with the matter fields
consisting of electrons, protons and neutrons on their path. Hence
the CP violation parameters will be modified due to such matter
effect as such interactions are not invariant under CP
transformation.  The general discussion of matter effect in the long
baseline experiments was given by Kuo and Pantaleone \cite{kuo}. The
T violation effects in the Earth were also studied numerically by
Krastev and Petcov \cite{petcov}. The data in the long baseline
experiments include the background matter effect which is not CP
invariant. Therefore, it is very important to investigate the matter
effect in order to estimate the CP violating effect originating from
the neutrino mixing matrix. The CP violation effect in long baseline
experiments are well studied in the literature
\cite{ref1,ref2,ref3}. Due to matter effect the evolution equation
becomes \cite{ref3}\be
 i \frac{d \nu}{dx}= {\cal H} \nu
\ee where \be {\cal H} \equiv \frac{1}{2E} \left ( U_m ~ {\rm
diag}(\mu_1^2, \mu_2^2, \mu_3^2)~ U_m^\dagger \right ). \ee The
matrix $U_m$ and the masses $\mu_i$'s are
determined by \bea U_m \left ( \begin{matrix} \mu_1^2 & 0 & 0\\ 0& \mu_2^2& 0\\
0&0 & \mu_3^2\\ \end{matrix} \right ) U_m^\dagger= U\left ( \begin{matrix} 0 & 0 & 0\\ 0& \Delta m_{21}^2& 0\\
0&0 & \Delta m_{31}^2\\ \end{matrix} \right ) U^\dagger
+ \left ( \begin{matrix} A & 0 & 0\\ 0& 0& 0\\
0&0 & 0\\ \end{matrix} \right ), \eea where $A= 2 \sqrt 2 G_F N_e E
= 7.56 \times 10^{-5} ~{\rm eV^2} \displaystyle{\frac{\rho}{\rm g~
cm^{-3}}\frac{E}{\rm GeV}} $ with $N_e$ is the electron density and
$\rho$ is the matter density. The solution of the above equation is
given as
$$ \nu(x) = S(x) \nu(0)
$$ with $$S= T e^{\int_0^x ds {\cal H}(s)},$$ giving the oscillation
probability for $\nu_\alpha \to \nu_{\beta},~ (\alpha, \beta = e,
\mu, \tau)$ at distance $L$ as \be P(\nu_\alpha \to \alpha_\beta;
L)=|S_{\beta \alpha}(L)|^2\;. \ee Thus, one can obtain a simple
approximative result for the appearance probability $P(\nu_e \to
\nu_\mu)$   as \cite{lind,lind1} \bea P(\nu_e \to \nu_\mu) & \approx
& \sin^2 \theta_{23} \sin^2 2 \theta_{13} \frac{\sin^2 [(\hat
A-1)\Delta_{31}]}{(1-\hat A)^2}\nn\\
& + & \alpha \sin \delta \cos \theta_{13} \sin 2 \theta_{12} \sin 2
\theta_{13} \sin 2 \theta_{23} \sin \Delta_{31} \frac{\sin (\hat A
\Delta_{31})}{\hat A} \frac{\sin[(1- \hat A) \Delta_{31}]}{(1-\hat
A)}\nn\\
& + & \alpha \cos \delta \cos \theta_{13} \sin 2 \theta_{12} \sin 2
\theta_{13} \sin 2 \theta_{23} \sin \Delta_{31} \frac{\sin (\hat A
\Delta_{31})}{\hat A} \frac{\sin[(1- \hat A) \Delta_{31}]}{(1-\hat
A)}\nn\\
&+& \alpha^2 \cos^2 \theta_{23} \sin^2 2 \theta_{12}
\frac{\sin^2(\hat A \Delta_{31})}{\hat A^2}, \label{eq2}
 \eea
 where $\alpha= \Delta m_{21}^2/\Delta m_{31}^2$, $\hat A= A/\Delta
 m_{31}^2$, and $\Delta_{31}=\Delta m_{31}^2 L/4E$. The first and
 last terms in eq (\ref{eq2}) correspond to the atmospheric and solar
 probabilities while the terms proportional to $\alpha$ are the
 interference between the solar and atmospheric contributions.

 A particularly interesting situation occurs for the case when
 \be \sin(\hat A \Delta_{31})=0, \Rightarrow \hat A \Delta_{31}=\pi, \label{eq3}\ee
 for which the $\delta $ dependence disappears  in the
 transition probability $P(\nu_e \to
\nu_\mu)$ as seen from Eq. (\ref{eq2}). This condition can also be
translated as \be \sqrt 2 G_F N_e L = 2 \pi ,\label{eq4} \ee which
is independent of the energy $E$.
 The baseline for which condition (\ref{eq4}) is
satisfied is known as magic baseline, which is basically found to be
\be \left (\frac{\rho}{\rm g/cc}\right ) \left ( \frac{L}{\rm km}
\right ) \simeq 32725. \ee This magic baseline is found to be  \be
L_{magic}=7690,\ee according to Preliminary Reference Earth Model
(PREM) \cite{prem} density profile of the earth. The implications of
such magic baseline is studied for the clean determination of
$\theta_{13}$ and $sgn(\Delta m_{31}^2)$ \cite{magic}. However, here
we are interested to see whether CP violation could be observed in
such magic baseline experiments.

Since the transition probability is independent of the CP violating
phase $\delta$ for $L_{magic}$, it is naively expected that CP
violation would also vanish for such experiments, however in actual
practice it is not the case. The intrinsic CP violation due to the
complex phase in the PMNS matrix which is proportional to $\sin
\delta$ vanishes whereas significant CP violation due to matter
effect could be possible.

 The transition
probability for $\bar \nu_\alpha \to {\bar \nu_\beta} $ can be
obtained from (\ref{eq2}) by replacing $\hat A \to -\hat A$ and
$\delta \to -\delta $. Thus the CP violating parameter in the
presence of matter can be given as \bea \Delta P (\nu_e \to \nu_\mu)
&\equiv & P(\nu_e \to \nu_\mu;L) - P
({\overline{\nu_e}}-{\overline{\nu_\mu}};L). \label{cpv1}\eea After
obtaining the relevant expressions for CP violation, we now proceed
to estimate its value both in the vacuum oscillation case
(\ref{cpv}) and including the matter effect contributions
(\ref{cpv1}). For numerical estimation, we use the central values of
the mixing angles and mass square differences as given in
(\ref{data}) and the baseline length as $L=7152$ km. Since the Dirac
CP violating phase $\delta$ is unconstrained, we vary its value
between $(10 - 90)^\circ$.  With these inputs, in Figure-1 we show
the variation of CP violation parameter (in vacuum) with beam
energy. From the figure it can be seen that CP violating effect of
few percent could be possible for such a long baseline experiment
and its dependence on the CP violating phase $\delta$ is quite
significant. In this case we get the same behavior for the CP
violating observable both in the normal as well as inverted
hierarchy cases of neutrino masses.

\begin{figure}[htb]
\centerline{\epsfysize 3.0 truein \epsfbox{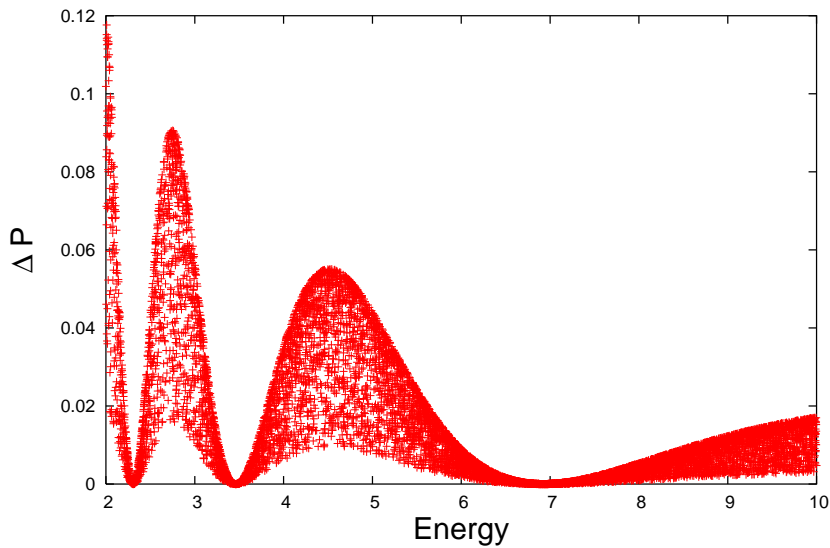}} \caption{The
variation of CP violating parameter (\ref{cpv}) with beam energy (in
GeV), where we have varied the CP violating phase between
$(10-90)^\circ$.}.
\end{figure}

 The corresponding variation,
including the matter effect (\ref{cpv1}) is shown in Figure-2 both
for normal hierarchy (red region) and inverted hierarchy (blue
region), where we have used the same input parameters as figure-1
and vary the CP violating phase $ \delta$ between $(10-90)^\circ $.
From the figure it can be also be noted that the dependence $\delta$
is almost negligible for such a baseline length. So the measurement
of CP violation in such experiment will also provide additional
information regarding the hierarchical nature of neutrino masses.

\begin{figure}[htb]
\centerline{\epsfysize 3.0 truein \epsfbox{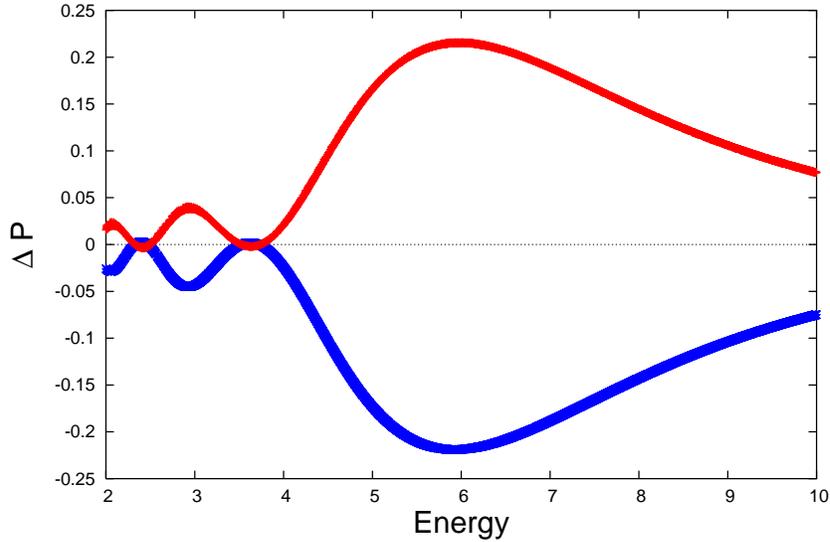}} \caption{The
variation of CP violating parameter including matter effect
(\ref{cpv1}) with beam energy (in GeV), where the red (blue) plots
correspond to normal (inverted) hierarchical behavior of neutrino
masses.}
\end{figure}

In figure-3 we have shown the CP violation effect (with normal
hierarchy)  for two representative baseline lengths : $L=2500$ km
and $L=5000$ km. In this case the $\delta$ dependence is not
completely negligible. For inverted hierarchy case the CP violation
effect will be opposite to that of normal case.

\begin{figure}[htb]
\centerline{\epsfysize 3.0 truein \epsfbox{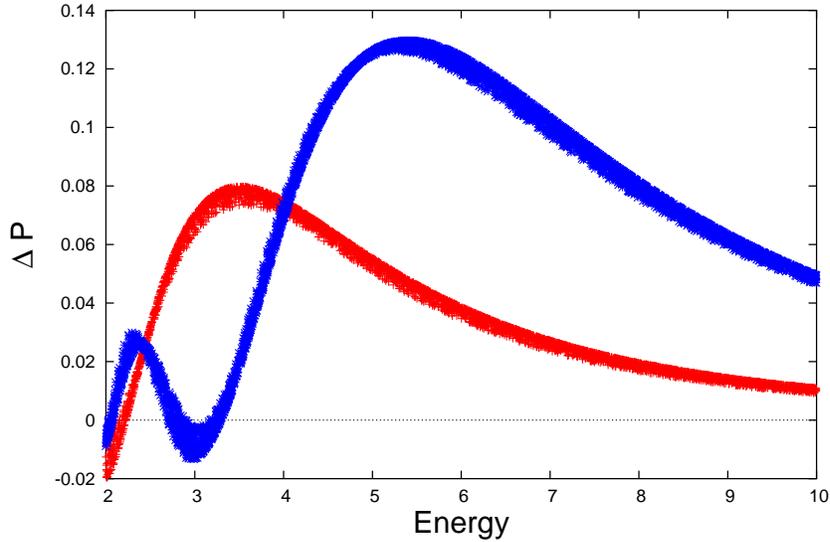}} \caption{Same
as Figure-2 with normal hierarchy for two different baseline
lengths, where the red (blue) regions are for $L= 2500 ~(5000)$ km.}
\end{figure}


To summarize, in this paper we have examined the possibility of
observing CP violation in the lepton sector in the proposed INO
experiment, using the beta beam from CERN.  In the lepton sector
also CP violation is expected unless neutrinos are exactly massless.
In particular CP violation in neutrino flavour oscillation is an
important phenomenon because it is  directly related to the CP
violating phase parameter in the mixing matrix. Unfortunately this
CP violating effect is suppressed in the short baseline accelerator
experiments if the neutrinos have hierarchical mass spectrum.
However the suppression is avoidable in the long baseline
accelerator experiments, which are expected to operate in the near
future. So there is probability that one can observe CP violating
effect in those experiments. We found that CP violating effect of
few percent could be observable at the INO detector using the beta
beam from CERN. We have also investigated the matter effect on the
CP violation parameter and found that it has significant
contribution for such base line length. We have shown that CP
violation effect as large as $\sim 20\%$ could be possible in such
experiment. Furthermore of CP violation in this experiment can also
provide us the evidence whether the neutrino masses are normal or
inverted hierarchical in nature. It is therefore strongly argued to
look for leptonic CP violation effect at INO.

\acknowledgements The work of RM was partly supported by Department
of Science and Technology, Government of India through grant no.
SR/S2/RFPS-03/2006.

\end{document}